\documentclass[%
reprint, superscriptaddress,
preprintnumbers,
amsmath,amssymb, aps,
twcolumn, showkeys]{revtex4-2}

\usepackage{graphicx}
\usepackage{dcolumn}
\usepackage{bm}
\usepackage{color}
\usepackage{hyperref}
\usepackage{float}
\usepackage{enumitem}
\usepackage{placeins}
\usepackage{lmodern}
\usepackage[dvipsnames]{xcolor}
\usepackage{comment}
\setlist[itemize]{leftmargin=*, label=\raisebox{0.25ex}{\tiny$\bullet$}, labelsep=0.5em, itemsep=0pt, topsep=0pt, parsep=0pt}

\usepackage{booktabs}
\usepackage{tabularx} 

\usepackage{lipsum}
\begin{document}


\title{The dynamics of discovery and the Heaps-Zipf relationship}

\author{Célestin Zimmerlin}
\affiliation{%
  CNRS, UMR Géographie-cités \& LabCom MIXTAPES\\
  Campus Condorcet\\
  FR-93322 Aubervilliers Cedex
}%

\author{Thomas Louail}
\affiliation{%
  CNRS, UMR Géographie-cités \& LabCom MIXTAPES\\
  Campus Condorcet\\
  FR-93322 Aubervilliers Cedex
}%
 
\author{Manuel Moussallam}
\affiliation{%
  Deezer research \& LabCom MIXTAPES
}%

\author{Marc Barthelemy}%
\affiliation{Université Paris-Saclay, CNRS, CEA, Institut de
   Physique Théorique, 91191 Gif-sur-Yvette, France}
 \affiliation{Centre d'Analyse et de Math\'ematique Sociales (CNRS/EHESS) 54 Avenue de Raspail, 75006 Paris, France}

\date{\today}

\begin{abstract}

  When following a sequence — such as reading a text or tracking a user’s activity — one can measure how the `dictionary' of distinct elements (types) grows with the number of observations (tokens). When this growth follows a power law, it is referred to as Heaps’ law, a regularity often associated with Zipf’s law and frequently used to characterize human 
  discovery processes. While random
  sampling from a Zipf-like distribution can reproduce Heaps’ law, this connection relies on the assumption of temporal independence — an assumption often violated in real-world systems although frequently found in the literature. Here, we investigate how temporal correlations in token sequences affect the type–token curve. In 
  human behaviors like music listening and web browsing, domain-specific correlations in token ordering lead to systematic deviations from the Zipf–Heaps framework, effectively decoupling the type–token plot from the rank–frequency distribution.
  Using a minimal one-parameter model, we reproduce a wide variety of type–token
  trajectories, including the extremal cases that bound all possible behaviors compatible with a given frequency distribution. Our results demonstrate that type–token growth reflects not only the empirical distribution of type frequencies, but also the domain-specific, temporal structure of the sequence — a factor often overlooked in empirical applications of scaling laws to characterize human behavior.

\end{abstract}

\preprint{Accepted in \textit{Physical Review E} \href{https://doi.org/10.1103/543d-frbq}{DOI: 10.1103/543d-frbq}}

\maketitle

\section{Introduction}
\label{sec:intro}

When observing a sequence—such as reading a text, browsing websites, or listening to music— the basic elements that make up the sequence (words, web pages, tracks) may either recur or appear for the first time, reflecting a continuous interplay between familiar elements and novel ones. Two empirical laws—\emph{Heaps’ law} and \emph{Zipf’s law}—have emerged as central tools for describing how novelty and frequency are distributed in these systems \cite{altmann_statistical_2024}. Heaps’ law characterizes the growth of the number of distinct types $D$ with the number of observed tokens $k$ in a sequence, typically following a sublinear power law.
\begin{equation}
    D \propto k^{\alpha},
    \label{eq:Heaps}
\end{equation}
with $\alpha \in [0.4, 0.7]$ in most empirical
cases~\cite{herdan_type-token_1960, heaps_information_1980}. It
has been observed in systems ranging from natural language and source
code to scientific and chemical databases~\cite{gelbukh_zipf_2001,
  zhang_discovering_2009, benz_discovery_2008}, and has been considered as a fitting law for capturing innovation, novelties and discovery processes.~\cite{tria_zipfs_2018,tria_dynamics_2014,di_bona_dynamics_2025}.

Zipf’s law \cite{zipf_human_1950}, in contrast, describes the distribution of type
frequencies. When types are ranked by decreasing frequency, the
frequency $f(r)$ of the type at rank $r$ follows
\begin{equation}
  f(r) \propto \frac{1}{r^{\nu}},
  \label{eq:Zipf}
\end{equation}
This pattern appears in diverse domains \cite{piantadosi_zipfs_2014} :
historically in the case of city sizes~\cite{auerbach_law_2023}, word
frequencies in texts~\cite{zipf_human_1950}, genome expression~\cite{furusawa_zipfs_2003}, and web page popularity in online behavior~\cite{adamic_zipfs_2002}, or music content \cite{manaris_zipfs_2005, zanette_zipfs_2006, serra_measuring_2012}.
A related formulation considers frequency $f$ as a random variable, with distribution

\begin{equation}
    p(f) \propto f^{-\gamma},
    \label{eq:Proba}
\end{equation}
where $\gamma = 1 + 1/\nu$~\cite{li_zipfs_2002,adamic_zipf_2000, stanisz_complex_2024}. Although both formulations have been referred to as “Zipf’s law”, the term denotes the rank–frequency relation~\eqref{eq:Zipf}. As they are only asymptotically equivalent~\cite{corral_distinct_2020} and for low ranks $r$ (i.e., large frequencies $f$), care should be taken not to conflate the two.

Heaps’ and Zipf’s laws often co-occur, with a widely reported relation
between their exponents~\cite{serrano_modeling_2009,vanleijenhorst_formal_2005}
\begin{equation}
    \alpha = \frac{1}{\nu}
    \label{eq:relation_exposant}
\end{equation}

This coexistence has been reproduced by theoretical models ~\cite{newman_power_2005}, such as the Yule–Simon process~\cite{yule_iimathematical_1925, simon_class_1955}, and has even inspired frameworks specifically designed to account for both laws, notably the adjacent possible model for innovation~\cite{tria_dynamics_2014, tria_zipfs_2018}. 
Despite their frequent association, a key conceptual distinction is that Zipf’s law is static: it describes aggregate frequency distributions, irrespective of token order. Heaps’
law, however, is inherently dynamic: it captures how novelty accumulates
over time. Any analytic link between the two thus relies on the strong
assumption of temporal independence, which is often violated in real-world systems.

Interestingly, in the case in written texts Heap's law can still be observed even though this assumption is known to be false. Despite known burstiness and thematic recurrence, random sampling
from empirical word-frequency distributions still reproduces reasonable
type–token curves~\cite{serrano_modeling_2009, fontanelli_quadratic_2025}. This has been attributed to the fact that temporal correlations affect the recurrence of known words more than the introduction of new
ones~\cite{font-clos_log-log_2015}.
Outside human language, however, these assumptions may not hold. In
digital environments --- such as streaming platforms, or even the world
wide web --- individual users select content from vast, ever-growing
catalogues, and their sampling strategies are not constrained by
syntactic or semantic rules. While the analogy with natural language is
tempting, it is reasonable to assume that the dynamics of exploration
will be different. Nevertheless, plotting a type–token curve for a
user’s sequence and fitting a Heaps-like law is straightforward. It
results in an exponent value $\alpha$ that it is convenient to use as
proxy for the user's ``average discovery rate'', with $\alpha < 1$
indicating sublinear growth (i.e., novelty 
declines with $k$ as $dD/dk \sim k^{\alpha - 1}$).


Numerous studies have questioned the validity of both Heaps'~\cite{gerlach_stochastic_2013,font-clos_log-log_2015,verbavatz_growth_2020} and Zipf's laws \cite{moreno-sanchez_large-scale_2016, gerlach_stochastic_2013,gan_is_2006,arshad_zipfs_2018, baayen_word_2003, piantadosi_zipfs_2014} across various systems for which they were previously applied. Moreover, Font-Clos and Corral~\cite{font-clos_log-log_2015,corral_distinct_2020} have questioned the very use of the rank–frequency plot, arguing that rank-based distributions $p(r)$ poorly capture low-frequency types and recommending instead the use of the frequency histogram $p(f)$. More recently, Fontanelli and Li~\cite{li_range-limited_2024} suggested that a concave type–token curve can emerge regardless of the underlying $p(r)$, implying that Zipf's law may not be mandatory for Heaps' law to hold. This hypothesis was already proposed in~\cite{vanleijenhorst_formal_2005} and remains, to our knowledge, an open question in the literature, leaving the equivalence between Heaps' law and Zipf's law under random sampling formally unestablished. Furthermore, while random sampling of a Zipf-distributed vocabulary is known to imply Heaps' law~\cite{vanleijenhorst_formal_2005}, several studies have proposed corrections to this relationship~\cite{lu_zipfs_2010,fontanelli_quadratic_2025,font-clos_log-log_2015}, using the alternative formulation of Zipf's law (see Eq.~\ref{eq:Proba})~\cite{font-clos_log-log_2015} or by incorporating finite-size effects~\cite{lu_zipfs_2010}.


In this paper, we do not engage in discussions on the Heaps–Zipf framework or its analytical consistency. Instead, we refer to studies that use these tools, and particularly Heaps’ exponent, as a measurement device drawing an analogy with written texts. ~\cite{di_bona_dynamics_2025, tria_dynamics_2014, bona_social_2022, tria_zipfs_2018, serra-peralta_heaps_2021, krawczyk_recovering_2023, abe_scaling_2021, li_range-limited_2024}. Many empirical datasets display Zipf-like rank–frequency distributions and Heaps-like concave growth, which are often fitted to obtain exponents $\gamma$ and $\alpha$, even though rigorous tests such as \cite{moreno-sanchez_large-scale_2016} would possibly fail. A remaining unresolved question is, therefore: what does this exponent actually measure ? Under random sampling, this exponent depends solely on $p(r)$, but what happens for real-world data, where the appearance of new types may be temporally or contextually correlated ? To what extent can such correlations affect the estimated value of Heaps’ exponent ?

In the following of this paper we aim to test the robustness of Heaps exponent's under real-world human discovery
dynamics, by comparing empirical sequences of text, individual music
listening histories and web browsing logs with their reshuffled counterparts, where
temporal correlations are removed. By estimating the scaling exponent
$\alpha$ in both real and reshuffled versions, we isolate the impact of
temporal correlations.
We will show that while written texts statistics are
largely robust to reshuffling, digital exploration sequences are not,
highlighting the role of system-specific temporal dynamics in shaping
discovery processes.
On the top row of Figure~\ref{fig:exponents_distribution} we provide an illustrative example of type–token curves for individual sequences taken from each of the three datasets used in the following of the paper. Corresponding rank-frequency plots are available in the Supplementary Information.

\section{Influence of temporal correlations on Heaps' law}

We compare here datasets documenting very different processes: individual listening histories on a music streaming platform; individual web browsing histories; and written texts from English language literature. Despite their differences, these systems have all been previously analyzed as instances of general processes of human exploration, discovery, or innovation~\cite{cherifi_returners_2016, tria_dynamics_2014, di_bona_dynamics_2025, louail_headphones_2017}. In particular, for music listening and written text, these dynamics have been quantitatively described within the Heaps’ law framework, where the scaling exponent $\alpha$ of the type–token relation (see Eq.~\ref{eq:Heaps}) is interpreted as a quantitative signature of the underlying discovery process.

More specifically, we contrast the classical case of vocabulary growth
in written texts (English literary classics) with that of individual
users who as time goes by listen to new tracks on a music streaming
platform over the course of 5 years --- that is, we study the growth of
their personal musical `vocabulary'. Following previous work ~\cite{di_bona_dynamics_2025, moscati_familiarizing_2025}, each individual track is considered a distinct type, meaning that discovery is defined as the first appearance of a given track in an individual user’s listening history, 
regardless of the artist, genre, or album they come from
(see \textit{Appendix A} for details on datasets). In addition, we analyze a complementary dataset of digital exploration consisting in individual web browsing histories collected over a month~\cite{kulshrestha_web_2021}. In this case a type is considered to be a domain name, and we ignore subdomains or sections within a website, i.e. browsing two distinct pages from the same website are here considered as two tokens but one and only type.

Since 
we are interested in modeling individual trajectories, we therefore characterize discovery as the first encounter of an item (or type) for each individual separately, regardless of others.  
The role and meaning of these discoveries for individuals, as well as the concrete opportunities for discovery associated with the "novelties" (i.e., the appearance of new types~\cite{tria_dynamics_2014}, which represent potential discoveries for each individual), vary across the different datasets. We return to these points in the following section, as they provide interpretation perspectives.

\begin{figure*}
  \hspace{-10mm}
  \centering
  \includegraphics[width=0.9\linewidth]{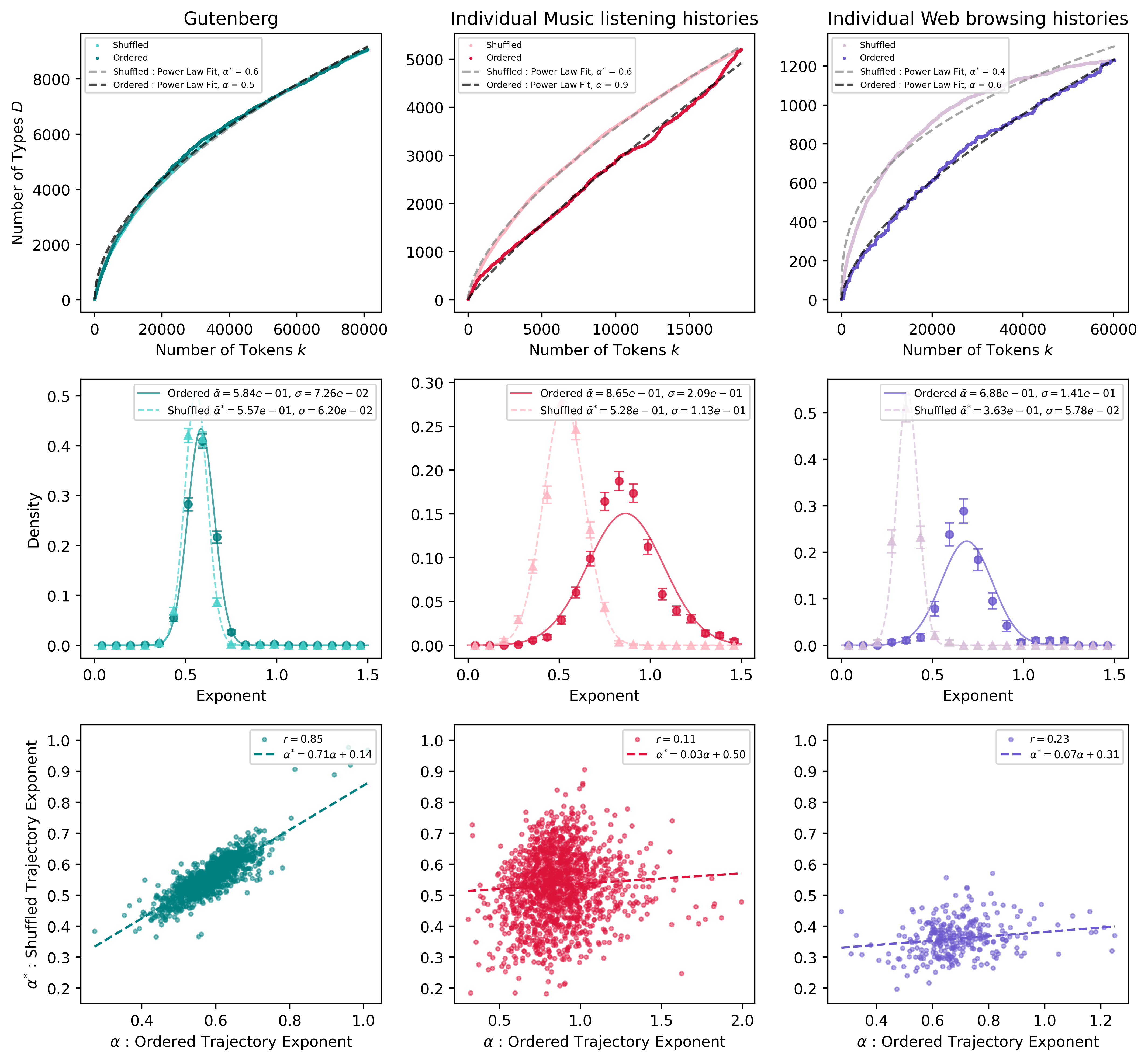}
    \caption{\footnotesize \textbf{Top:} Type–token plots for a random sample from each dataset: texts from the Project Gutenberg corpus (left), individual music listening histories of Deezer users (middle),and individual web browsing histories (right). For each case, both the empirical ordered trajectory and a reshuffled trajectory are shown. The scaling relation $D = c k^\alpha$ (\ref{eq:Heaps}) is fitted using two alternative methods; for each sequence, the exponent corresponding to the fit with the lowest Bayesian Information Criterion (BIC) is retained (see Appendix B).
    \textbf{Middle:} Distributions of scaling exponents $\alpha$ obtained for both ordered and reshuffled sequences $D(k)$, across the three datasets. We consider more than 1000 sequences for both the Gutenberg and Deezer datasets, and 292 sequences for the Web
    Tracking dataset. Error bars are computed using the bootstrap method
    described in~\cite{sohil_introduction_2022}, with 1,000 bootstrap
    samples. \textbf{Bottom:} Relationship between exponent values
    estimated for the ordered ($\alpha$) and reshuffled ($\alpha^{*}$)
    sequences of the same individual user or text. Each plot reports the Bravais–Pearson
    correlation coefficient $r$, along with a linear regression
    $\alpha^{*} = a \alpha + b$ fitted by minimizing the absolute
    deviation to reduce sensitivity to outliers.}
    \label{fig:exponents_distribution}
\end{figure*}

For these three case studies, we compute the scaling exponents obtained by
fitting Heaps' law (see Eq.~\eqref{eq:Heaps}) to both the original token
discovery sequences and their reshuffled counterparts. The reshuffled
sequences correspond to a scenario in which each token is drawn
independently at random from the empirical rank–frequency distribution
$p(r)$. The second row of Figure~\ref{fig:exponents_distribution} shows the distribution of the exponent values obtained for both cases—ordered and reshuffled sequences—across the different datasets. We denote by $\alpha$ the scaling exponent estimated from the original (ordered) sequence, and by $\alpha^*$ the exponent obtained from the reshuffled sequence

For the case of texts (left panel of Figure~\ref{fig:exponents_distribution}), we see on the bottom line that the exponent values computed from the observed (ordered) sequences are highly correlated with those obtained from reshuffled sequences. The overall distributions of exponent values are also very similar in the two cases (center line), indicating that the temporal structure of token occurrences has little effect on the scaling behavior. In contrast, individual music listening histories over several years show a markedly different pattern. The scaling exponents obtained from the observed sequences are, on average, significantly higher than those obtained from reshuffled data, and they exhibit greater variability (i.e., larger variance). Moreover, the correlation between the real and reshuffled exponent values is weak, as shown in the bottom row of the figure. Finally, the exponents obtained for web browsing trajectories display a similar pattern to those obtained for music listening : the exponent values derived from the observed sequences $D(k)$ are consistently larger than those from reshuffled sequences. Here too, the correlation between the two sets of exponents is weak (bottom row),
although the effect is somewhat less pronounced than in the music case. 
We compare the $R^2$ scores of power-law fits for ordered and reshuffled trajectories across datasets (see Fig.~\ref{fig:R_squared}). Rather than assessing statistical significance, this measure here serves us as a simple tool to compare the power-law fits of individual trajectories, both ordered and reshuffled, in each dataset and across datasets. Overall, the $R^2$ values are very high ($\bar{R^2} \approx 0.99$) (see Appendix B for further details).

\medskip

These observations suggest that the scaling exponent $\alpha$ obtained
by fitting Heaps' law (Eq.~\ref{eq:Heaps}) to individual exploration
sequences $D(k)$ in web browsing or music listening is shaped not only
by the empirical rank–frequency distribution $p(r)$ (which reflects how
the individual's attention is distributed among items), but also by the
underlying temporal dynamics. This is in contrast to vocabulary growth
in texts, where the exponent obtained from the empirical sequence can be
accurately reproduced by randomly sampling tokens from $p(r)$. In particular, in the case of musical exploration, the exponent values
obtained for the ordered case (observed discovery sequences) are much
closer to $\alpha = 1$ than the ones obtained when reshuffling the same
music listening sequences (for which we obtain $\bar{\alpha^*}=0.5$). $\alpha=1$ corresponds to the linear case of a stationary integration of discoveries into the listeners's catalogue. Interestingly, a substantial portion of our sample also exhibits exponents values $\alpha > 1$, which appears to contradict the standard formulation of Heaps’ law \cite{chacoma_heaps_2020,lu_zipfs_2010}. Heaps’ law is traditionally understood as an asymptotic result, and in the asymptotic regime it is
indeed true that due to the constraint $D < k$, then necessarily $\alpha < 1$. However, in a more general framework applied to finite-length  sequences, values of $\alpha > 1$ are entirely possible, as long as the prefactor $c$ in the relation $D = ck^\alpha$ remains sufficiently small (in this case $ck^\alpha<k$ implies that $k<k_c$ with $k_c=1/c^{1/(\alpha-1)}$). These $\alpha > 1$ cases correspond to
situations in which the appetite of the individual for discovery grows
over time. It is worth noting that some prior works on individual music discovery enforced fitting procedures that systematically prevent the emergence of exponents $\alpha > 1$~\cite{di_bona_dynamics_2025,bona_social_2022}, potentially overlooking this important dynamical feature. The observation of these super-linear exploration trajectories in musical contexts, as opposed to written texts, points to a fundamental difference between the catalogues under study. On streaming platforms and the web, the set of available items is not only larger than in written corpora, but also expands over time at a rate that greatly exceeds the pace of individual human exploration of these digital archives. This fundamental constraint helps explain why the space of discovery trajectories differs between systems, simply because the possibility of continuing to discover through actual novelties depends on the nature of each system.

\medskip

Although the effects of correlations in music and web browsing discovery trajectories manifest similarly in the reshuffling experiment associated with Fig.~\ref{fig:exponents_distribution}, this does not imply that they are of the same nature, nor driven by the same processes. To get a first insight on the role of memory and temporal direction in shaping correlations, we reproduce the plots of Fig.~\ref{fig:exponents_distribution}, but this time by using time-reversed trajectories --- that is, the original sequences with their temporal order inverted. The results are heterogeneous across datasets, highlighting that temporal correlations, and their manifestations in the type-token plots, can take diverse forms depending on the system. Interestingly 
for the case of music listening the exponent values of the time-reversed trajectories are much closer to the values obtained for the reshuffled trajectories than those obtained for the ordered ones (see Fig.~2 in Supplementary Information). 
We note that this is not the case for the two other systems: for books the time-reversed, ordered and reshuffled trajectories are all alike, and for web browsing the time-reversed sequences resemble the ordered ones. This observation raises interesting questions about the role of memory in discovery processes, which would deserve further study.

\section{Dependence of scaling deviations on sequence length}

\begin{figure*}
  \centering \includegraphics[width=0.99\linewidth]{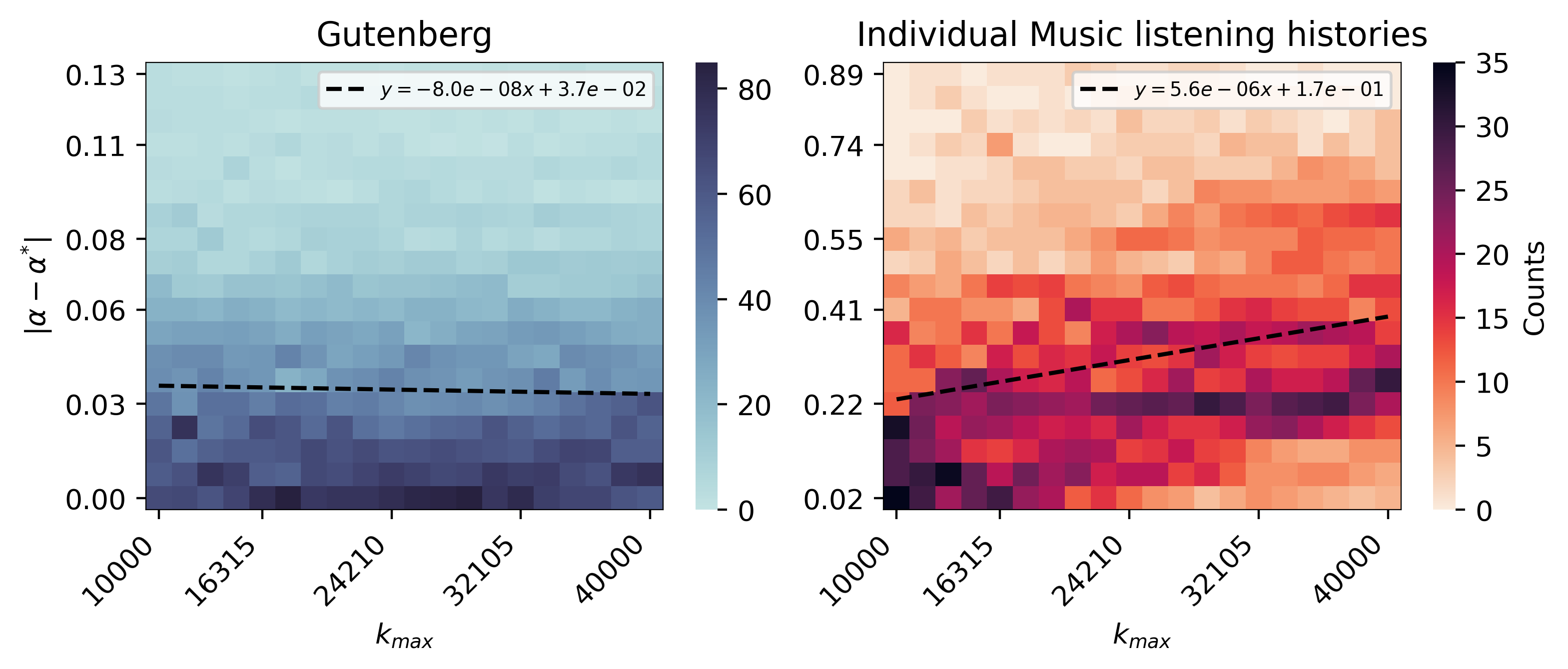}
  \caption{Evolution of the absolute difference $|\alpha - \alpha^*|$ as
    a function of the sequence length $k_{\max}$. Each sequence (either
    a text or an individual music listening history) is truncated at 20
    different values of $k_{\max}$, ranging from 10{,}000 to 40{,}000
    tokens. For each truncation point, we compute the absolute
    difference between the exponent obtained for the ordered sequence
    ($\alpha $) and the one obtained for the reshuffled sequence
    ($\alpha^*$). 
    The resulting values of $|\alpha - \alpha^*|$ are displayed as a heatmap. To highlight the overall trend, we further compute, for each $k_{\max}$, the average $|\alpha - \alpha^*|$ across all sequences and fit a linear regression to these averages.
    The analysis is based on 500 texts and 200 individual listening
    histories. The web browsing dataset is excluded due to insufficient
    sequence length (see Appendix A).}
  \label{fig:delta_exp_seq_length}
\end{figure*}

We have shown that individual music listening (over the course of
several years) and web browsing (over several weeks) exhibit exploration
dynamics that differ markedly from the paradigmatic case of written
texts, where Heaps' law was originally observed. In the latter, the
accumulation of novel elements follows a relatively smooth and universal
trend, whereas in music or web navigation, discovery is intertwined with
strong repetition and listener-specific content preferences \cite{sguerra_ex2vec_2023,sguerra_discovery_2022}. One important
consequence of these differences is that, for a given empirical
rank–frequency distribution $p(r)$—which gives the number of tokens
associated with each type—very different scaling exponents $\alpha$ can
emerge depending on the discovery process, that is on the temporal
ordering of the token sequence $D(k)$. The bottom line of Figure~\ref{fig:delta_exp_seq_length} shows that the exponent $\alpha$ measured on empirical (ordered) sequences is often substantially different from the value $\alpha^*$ obtained by
reshuffling the same sequence, which corresponds to a scenario of random
exploration drawn from the same catalog. This discrepancy indicates that
the exponent of the power-law fit reflects not just the frequency distribution $p(r)$, but
also system-specific temporal correlations in the discovery process.

To better understand this deviation, we investigate how the absolute
difference $|\alpha - \alpha^*|$ depends on the length of the
sequence. To do this, we truncate each sequence at different values of
$k_{\max}$, and compute the absolute difference
$|\alpha - \alpha^*|$ for each truncation
point. Figure~\ref{fig:delta_exp_seq_length} shows the results as a
heatmap (darker colors indicate higher point density), along with a
linear regression line to guide the eye. As shown in this figure, we find that for music listening, the difference $|\alpha - \alpha^*|$ increases with sequence length, unlike in the case of written texts where the difference is essentially small and constant. This result suggest that human exploration of large music
catalogs increasingly deviates from random exploration as the sequence
grows longer. It further indicates that estimating a single exponent
$\alpha$ over a given sequence compresses multiple heterogeneous factors
into a single parameter—namely, the attention distribution $p(r)$ and
the temporal correlations of the sequence (e.g., repetition vs. novelty)
compressed into the scaling exponent $\alpha$. The relative
contributions of these factors themselves depend on the sequence length,
which is often arbitrarily constrained by data availability.

\section{A toy model to explore the role of temporal correlations}

\subsection{The envelope of all possible Heaps curves}

\begin{figure}[ht]
    \centering
    \includegraphics[width =0.99\linewidth]{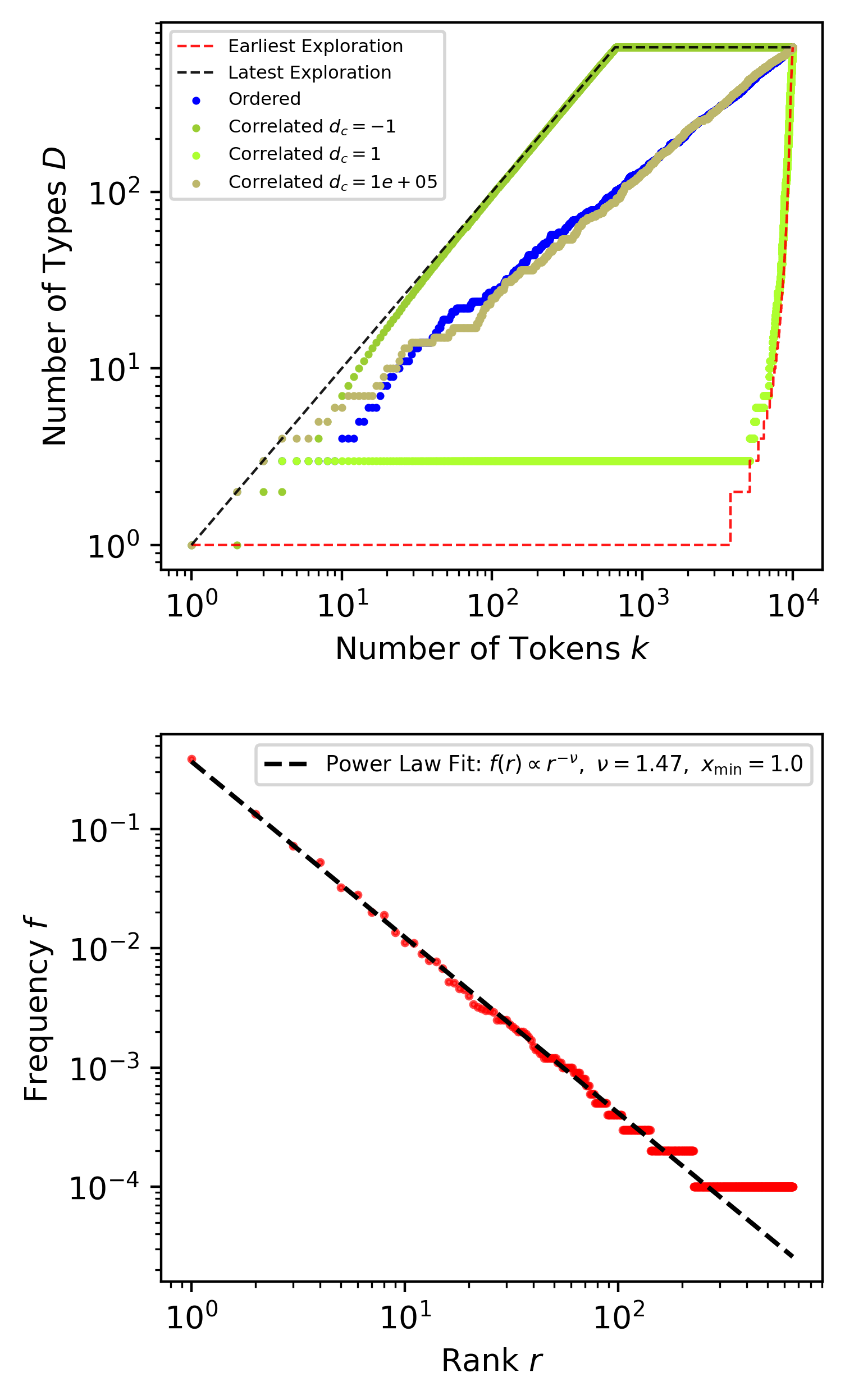}
    \caption{\textbf{Top:} Type–token trajectories for different
      orderings 
      of a given set of tokens,
      (sampled from $p(r) \propto r^{-1.5}$), generated by a simple toy model and showing variability in discovery dynamics for the same underlying distribution. The black and red dashed
      lines correspond to the limiting cases of maximally accelerated
      and delayed discoveries and bound all possible discovery trajectories.  \textbf{Bottom:} Corresponding rank–frequency distribution of types in the sequence, plotted in log–log scale.
      }
    \label{fig:Extremes_Trajectories}
\end{figure}

When the assumption of temporal independence is lifted and all possible
orderings of a sequence are allowed, the type–token trajectory $D(k)$
can vary widely—even for a fixed rank–frequency distribution $p(r)$. In
particular, two extreme cases can be identified that form the envelope
of all possible trajectories. These are shown as black and red dashed
lines in Figure~\ref{fig:Extremes_Trajectories} (top). They correspond
to `maximally accelerated' and `maximally delayed' discoveries of new
types, depending solely on the temporal ordering of the tokens. The black dashed line corresponds to the deterministic case where all $D_{\max}$
distinct types are discovered at the beginning of the sequence—one new
type is introduced at each step until $k = D_{\max}$. The remaining
$k_{\max} - D$ tokens are then repetitions of already observed
types. The other extreme case, shown in red, corresponds to a sequence where discoveries are delayed as much as possible while remaining
consistent with the frequency distribution $p(r)$. In this case, the
most frequent type (rank 1, with frequency $f(1) = k_1$) appears $k_1$
times at the beginning, followed by $k_2$ repetitions of the second most
frequent type, and so on, until the least frequent type is introduced at
the very end. This strategy delays the growth of the dictionary to reach
the final value $D$ only near $k = k_{\max}$. These two deterministic 
cases form the envelope of all possible type–token trajectories compatible
with a given $p(r)$.

The variability of these curves is illustrated in
Figure~\ref{fig:Extremes_Trajectories} using synthetic sequences
generated from a Zipf-like distribution of type frequencies
$p(r) \propto r^{-1.5}$, where different temporal orderings produce
widely varying $D(k)$ curves. The bottom panel shows the corresponding
rank–frequency distribution, fitted with a power-law using the method
of~\cite{clauset_power-law_2009}.

\subsection{A toy model}

This envelope of possible behaviors can be reproduced using a simple toy
model, controlled by a single parameter that modulates local temporal
correlations. We start from a synthetic set of tokens, sampled at random
from a rank–frequency distribution that follows Zipf's law (see
Eq.~\ref{eq:Zipf}). The purpose of the model is to iteratively build a
sequence from this set 
by using a stochastic mechanism that introduces temporal correlations
between tokens of the same type.

The general principle is the following: the sequence is generated one
step at a time, and at each step $k$, each token still available (i.e.,
not used yet) is selected with a probability that depends on the time
elapsed ('age') since the last appearance of its type in the sequence up to that point. For types that have not yet appeared in the sequence, we set the last appearance to step $0$, so that at step $k$ their age is $k$.

The probability of selecting a token of type $i$ at step $k$ is chosen
to be:
\begin{equation}
  p(i) \propto z(i) \cdot \exp\left(-\frac{k - l(i)}{d_c}\right)
\end{equation}
where: $z(i)$ is the number of remaining tokens of type $i$, $l(i)$ is
the index of the last position of this type in the sequence, and $d_c$
is a correlation length parameter controlling the memory of the
process. By varying the value of $d_c$, the model can generate a wide
spectrum of temporal correlation regimes:

\begin{itemize}
\item{} When $d_c \to 0^+$, the model favors recently used types,
  leading to bursty dynamics and delayed discovery (stochastically approaching the maximally delayed, deterministic discovery strategy previously described).
\item{} When $d_c \to 0^-$, the model favors types that have rarely (or
  never) appeared, accelerating discoveries as much as possible (and tends towards the maximally accelerated case).
\item{} When $d_c \to \infty$, the memory vanishes and the sequence
  becomes an independent random sampling from $p(r)$, recovering the
  classical Heaps’ law.
\end{itemize}

Note that the limiting trajectories (dashed lines) result from a deterministic algorithm, whereas the correlated trajectories are generated through the realization of a statistical model, which accounts for the small discrepancies between the two. Moreover, our approach is related to an existing model \cite{cattuto_semiotic_2007}, which iteratively generates a sequence while introducing similar temporal correlations. That model captures both the content and temporal arrangement of the sequence, whereas our approach focuses on the temporal organization of a pre-existing set of elements. We do not model how content appears, but instead demonstrate that local temporal correlations alone can generate a wide variety of $D(k)$ trajectories, even when the underlying distribution $p(r)$ is held fixed.


\section{Discussion}
\label{sec:discussion}

We compared different systems --- written texts in English literature, individual music listening histories, and web browsing histories --- that can be represented as ordered sequences of discrete tokens taken from different types (words, songs or web pages). We compared these systems by fitting a power-law of the form $D = c k^\alpha$ to both the empirically observed (ordered) sequences $D(k)$ and to reshuffled versions of the same sequences. In written texts, our results indicate that the appearance dynamics of
new types (i.e., new words), as captured by the curve $D(k)$, can be
well-approximated by a random exploration of a fixed vocabulary. In this
case, the type--token trajectory depends primarily on the empirical
frequency distribution $p(r)$. This finding aligns with previous
analytical results showing that Heaps’ law can emerge from random
sampling over a Zipfian distribution~\cite{vanleijenhorst_formal_2005},
and supports the idea that such an approach offers a reasonable starting
point for modeling vocabulary growth in natural language. However, we showed that this assumption breaks down in the two other
systems we studied --- individual online music listening and web
browsing. In these cases, the individual discovery process --- and
consequently the interplay between repetitions of already known tracks
and pages, and the temporal correlations in the ordering of token
appearances --- play a central role in shaping the observed type--token
trajectories. To further illustrate this point, we introduced a minimal
toy model that generates correlated sequences by modifying the temporal
structure of token occurrences. Despite its simplicity, the model
reveals that local correlations can drastically reshape the type--token
plot. By tuning a single parameter controlling the strength of
correlations, the model reproduces a wide range of behaviors, including
the envelope that bounds all possible trajectories in the type--token
plane for a given set of tokens.

\medskip

While the frequency–rank plot is appropriate for a
`static' study of discovery phenomena --- as it captures the overall
composition of a sequence --- it does not convey the temporal
dynamics of discovery, which are instead reflected in the Type–Token
plot. Because these two objects are fundamentally different in nature,
attempting to formally relate them is only possible under very specific
conditions, which are generally not satisfied in real sequences. By
default, the scaling exponent value $\alpha$ obtained when fitting a
power-law on the Type-Token plot aggregates phenomena of different
natures: it depends both on the type-frequency distribution $p(r)$ and
on the system-specific temporal dynamics. Consequently, quantitative
comparisons of Heaps' law exponent values across different systems or different individuals  are of limited interpretative value. Moreover, we found that the relative impact of temporal correlations on the estimated exponent can itself depend on the length of the sequence, further complicating comparisons. More generally, applying the Heaps/Zipf framework across a wide variety of phenomena is risky and should be done without assuming that all associated properties hold, such as the relationship between exponents as in \eqref{eq:relation_exposant}. 
Beyond previously noted issues—such as the conceptual limitations of rank plots, the empirical fragility of Zipf’s and Heaps’ laws, and their instability under random sampling—we emphasize an additional point: temporal correlations in sequences can significantly influence the type–token curve.

\medskip

These correlations are rarely analyzed, yet ignoring them may lead to misattributing Heaps' exponent values to intrinsic content diversity rather than to exploration dynamics. A related point that is important to underline is that, contrary to what is often stated in the literature, the Heaps exponent should not be interpreted as a direct measure of a system’s or individual's propensity for discovery (referred to in previous studies as the 'discovery rate'). Instead, it characterizes the dynamics of discovery itself, such as the rate of growth, decay, or stationarity over the course of the sequence. The prefactor obtained in a power-law fit of the form eq.~\eqref{eq:Heaps} further quantifies the rate of discovery.

\medskip

All these considerations support using the Zipf–Heaps framework primarily as a descriptive lens. While useful for identifying regularities, it does not offer rigorous statistical signatures, let alone universal laws of innovation. Quantitative comparisons across systems of different nature should therefore be approached with caution. More broadly, the difficulty of reducing heterogeneous phenomena to a single Heaps exponent highlights a deeper issue: collapsing diverse modes of exploration into a universal model risks erasing crucial domain-specific differences, such as the influence and depth of memory. An alternative is to examine the interplay between exploration and repetition within specific contexts, which can reveal internal semantics and the particular ways people engage with different forms of content.


\section*{Acknowledgments}

This project has received financial support from the CNRS through the MITI interdisciplinary and exploratory research program. T.L., M.M., and M.B. supervised this work. C.Z. performed the data analysis, coding, and initial draft writing. All authors contributed to the study’s conceptualization, validation, final writing, review, and editing. C.Z thanks the Deezer Research team and the Sony Computer Science Laboratory (Paris) for useful comments and support, and Emile Emery for his feedback.

\bibliographystyle{apsrev4-2}
\bibliography{references}
\clearpage

\appendix
\onecolumngrid

\begin{table*}[h]
    \squeezetable
    \footnotesize
    \centering
    \caption{\label{tab:likelihoods} Fitting methods}
    \begin{ruledtabular}
    \begin{tabular}{l c l}
        \textrm{Noise type} & \textrm{Log-likelihood} & \textrm{Fitting technique} \\
        \colrule
        Additive $y_i=\hat{y}_i+\epsilon_i$
        & $\ln L(\hat{y},\hat{\sigma})= -\tfrac{n}{2}\ln(2\pi\hat{\sigma}^2 e)$
        & Least squares on $y_i$ \\
        Multiplicative $y_i=\hat{y}_i \exp(\epsilon_i)$
        & $\ln L(\hat{y},\hat{\sigma})= -\tfrac{n}{2}\ln(2\pi\hat{\sigma}^2 e)-\sum_{i=1}^{n}\ln y_i$
        & Least squares on $\ln y_i$ \\
    \end{tabular}
    \end{ruledtabular}
\end{table*}

\begin{figure*}[h]
  \centering
  \includegraphics[width=1\textwidth]{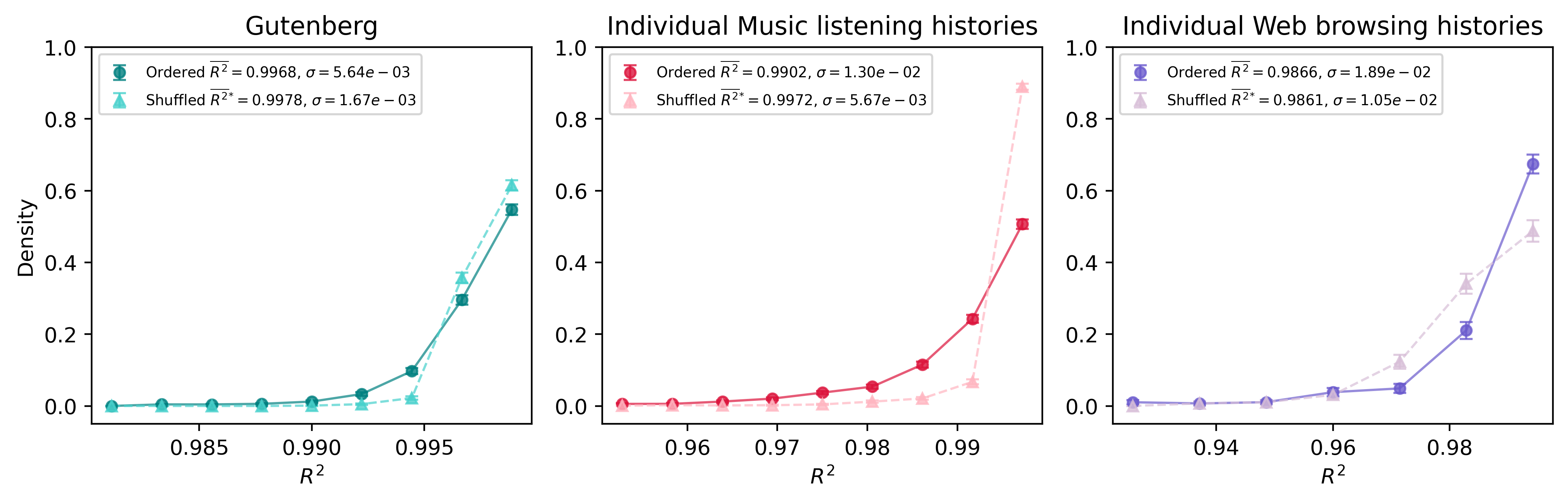}
      \caption{Distribution of the coefficient of determination ($R^2$) when fitting the power-law $D=ck^\alpha$ on the type-token plot (\ref{eq:Heaps}), for both ordered and reshuffled trajectories across the three datasets. Error bars are computed using a bootstrap procedure with $1000$ resamples, following the approach described in~\cite{sohil_introduction_2022}.}
  \label{fig:R_squared}
\end{figure*}
\twocolumngrid
\section{Datasets}

We use three different datasets covering textual, musical, and web
browsing activities:

\begin{itemize}
\item \textbf{Gutenberg.} We rely on the Project Gutenberg corpus,
  consisting of $1400$ texts listed in the catalog available
  \href{https://www.gutenberg.org/cache/epub/feeds/pg_catalog.csv.gz}{here}. From
  this set, we retain $1178$ texts selected such that $k_{\max} > 10^4$.

\item \textbf{Individual online music interaction data.} We analyze
  anonymized and timestamped listening histories of individual users of
  the streaming platform Deezer, in France, provided within the RECORDS project~\cite{renisio_integrating_2024}. A play is recorded in an individual’s listening history once the track has been continuously streamed for at least 30 seconds, in line with standard industry practices in particular for compensating rights holders. The data collected in this project include the listening history data on Deezer since 2018 of about 16,000 individuals in France, along with their responses to a detailed online survey
  about their listening habits, music preferences and cultural practices
  beyond music. The survey includes a standard socioeconomic module
  which allowed to collect precise information about the social
  characteristics of the survey participants, and consequently to analyze the social structure of contemporary music listening practices. The public release of several datasets combining listening data with survey data is planned for 2026.

  A subset of $3000$ listeners with complete listening histories over five
  years was first selected (2018 - 2023). Among these, $68\%$ meet the criterion
  $k_{\max} > 10^4$. From this subgroup, $1400$ listeners were sampled at
  random. 
  
  These individual listening histories are available from the corresponding author upon reasonable
  request.

\item \textbf{Web Tracking Dataset.} We use the dataset published by
  \cite{kulshrestha_web_2021}, available on
  \href{https://zenodo.org/records/4757574}{Zenodo}, which contains one
  month (October 2018) of web tracking data for $2148$ German individuals. We
  retain $292$ individuals with $k_{\max} > 9000$. This threshold is
  comparatively more restrictive: it corresponds to individuals who
  opened more than $9000$ web pages in a single month. This introduces a
  possible selection bias toward particularly active listeners. Furthermore,
  the relatively short sequence lengths limit the ability to study the
  longitudinal evolution of the gap between the fitted and reshuffled
  exponents (see ~\ref{fig:delta_exp_seq_length}).
\end{itemize}

\section{$R^2$ Comparison}

To evaluate the power-law fits, we compute the coefficient of determination ($R^2$) for all trajectories and plot their distributions in Fig ~\ref{fig:R_squared}.

We find excellent $R^2$ values, which should nonetheless be interpreted with caution: this indicator measures the correlation between model and data rather than the statistical significance of the fit, and it tends to be overly optimistic when evaluating power-law scaling~\cite{leitao_is_2016}. The purpose here is not to determine whether the data were truly generated by a power-law process—Eq.~\eqref{eq:Heaps} is clearly a simplification of real world complexity—but to use this metric to compare fit quality across different trajectories.

As shown in Figure \ref{fig:R_squared}, power-law fits to Eq.~\eqref{eq:Heaps} are generally better for text data than for music listening or web browsing sequences. For music and, to a lesser extent, texts, the real (ordered) sequences produce lower and more variable $R^2$ values compared with reshuffled sequences. In contrast, web browsing sequences show slightly higher average $R^2$ for ordered trajectories than for reshuffled ones, although the dispersion is larger.

Overall, this reinforces two key points:
1) Correlations in the temporal ordering of events can systematically affect the empirical fit to Heaps’ law.
2) This effect depends on the dataset considered, and is particularly pronounced in the case of music listening.

\section{Fitting Procedure}

We fit Heaps' law in type--token plots \eqref{eq:Heaps} using two
alternative assumptions about the nature of the noise:

\begin{enumerate}
\item \textbf{Additive noise.} This corresponds to minimizing the
  squared residuals between predictions and observations, i.e. using
  \texttt{scipy.curve\_fit} on the original scale.
\item \textbf{Multiplicative noise.} This corresponds to minimizing the
  squared residuals between the logarithm of predictions and the
  logarithm of observations, i.e. using \texttt{sm.OLS} on the
  log-transformed data.
\end{enumerate}

To guard against initialization effects, the first $1\%$ of each fitted trajectory is omitted.
For each trajectory, the retained exponent is the one obtained from the
fit with the lowest Bayesian Information Criterion (BIC). The
likelihoods corresponding to each noise assumption are summarized in
\autoref{tab:likelihoods}. This procedure was employed to determine the
exponents of the complete trajectories
(see \autoref{fig:exponents_distribution}). When examining the exponent
evolution with $k_{\max}$ (see \autoref{fig:delta_exp_seq_length}), we only used the additive noise hypothesis for convenience.


\end{document}